\documentclass[%
 reprint,
 superscriptaddress,
 amsmath,amssymb,,mathtools,lipsum
 aps,
 pra,
]{revtex4-2}
\usepackage{amsmath}
\usepackage[dvipsnames]{xcolor}
\usepackage{hyperref}
\usepackage{braket}
\usepackage{lineno}
\hypersetup{colorlinks=true,
            linkcolor=blue,
            urlcolor=blue,
            citecolor=blue}
\usepackage[skins,theorems]{tcolorbox}
\usepackage{sidecap}
\usepackage{wrapfig}
\usepackage{graphicx}
\usepackage{dcolumn}
\usepackage{bm}
\definecolor{lime}{HTML}{A6CE39}
\DeclareRobustCommand{\orcidicon}{%
	\begin{tikzpicture}
		\draw[lime, fill=lime] (0,0) 
		circle [radius=0.16] 
		node[white] {{\fontfamily{qag}\selectfont \tiny ID}};
		\draw[white, fill=white] (-0.0625,0.095) 
		circle [radius=0.007];
	\end{tikzpicture}
	\hspace{-2mm}
}

\foreach \x in {A, ..., Z}{%
	\expandafter\xdef\csname orcid\x\endcsname{\noexpand\href{https://orcid.org/\csname orcidauthor\x\endcsname}{\noexpand\orcidicon}}
}

\DeclareUnicodeCharacter{003BC}{}
\begin{document}
\preprint{APS/123-QED}

\title{Nondipole strong-field approximation for above threshold ionization in few-cycle pulse}

\author{\orcidA{}Danish Furekh Dar}
\email{danish.dar@uni-jena.de}
\affiliation{Helmholtz-Institut Jena, Fr\"o{}belstieg 3, D-07743 Jena, Germany}%
\affiliation{GSI Helmholtzzentrum f\"ur Schwerionenforschung GmbH, Planckstrasse 1, D-64291 Darmstadt, Germany}
\affiliation{Theoretisch-Physikalisches Institut, Friedrich-Schiller-Universit\"at Jena, Max-Wien-Platz 1, D-07743
Jena, Germany}

\author{\orcidB{}Bj\"orn Minneker}
\affiliation{Helmholtz-Institut Jena, Fr\"o{}belstieg 3, D-07743 Jena, Germany}%
\affiliation{GSI Helmholtzzentrum f\"ur Schwerionenforschung GmbH, Planckstrasse 1, D-64291 Darmstadt, Germany}
\affiliation{Theoretisch-Physikalisches Institut, Friedrich-Schiller-Universit\"at Jena, Max-Wien-Platz 1, D-07743
	Jena, Germany}

\author{\orcidC{}Stephan Fritzsche}
\affiliation{Helmholtz-Institut Jena, Fr\"o{}belstieg 3, D-07743 Jena, Germany}%
\affiliation{GSI Helmholtzzentrum f\"ur Schwerionenforschung GmbH, Planckstrasse 1, D-64291 Darmstadt, Germany}
\affiliation{Theoretisch-Physikalisches Institut, Friedrich-Schiller-Universit\"at Jena, Max-Wien-Platz 1, D-07743
Jena, Germany}

\date{\today}

\begin{abstract}
	The ionization of atoms and molecules by strong laser fields has been studied extensively, both theoretically and experimentally. The strong-field approximation (SFA) allows for the analytical solution of the Schr\"{o}dinger equation and accurately predicts the behavior of ionization processes in intense laser fields. Over the past decade, there has been a growing interest in the study of nondipole effects in these processes. However, such predictions have, so far, been limited to monochromatic driving laser fields, while experiments often employ quite short pulses. In this paper, we, therefore, present an extension of the SFA that also allows incorporating the more complicated temporal structure of a few-cycle pulse. By this extension, the prediction of co-called peak shifts is significantly improved, and the ability to control the laser pulse inducing above threshold ionization is greatly enhanced. The enhanced control over the characteristics of the laser pulse results in more accurate predictions of peak shifts. Our results show better agreement with experimental investigations compared to previous theoretical studies.
\end{abstract}
\maketitle

\section{Introduction}

The interaction of atoms and molecules with a high-intensity laser field has gained huge interest over the past few decades. When atoms are exposed to high-intensity laser fields, they exhibit complex behavior. These interactions are studied to better understand the fundamental properties of atoms, as well as to explore potential applications in fields such as spectroscopy and laser-based technologies. The dynamics of an electron in such an intense laser field have been characterized by several processes, including above threshold ionization \cite{agostini,paulus,boning1}, high harmonic generation \cite{ferray1988,mcpherson1987,Paufler} and non-sequential double ionization \cite{Ihuillier1983,Lhuillier1983,watson1997,Fang}.

The time-dependent Schr\"{o}dinger equation (TDSE) is the fundamental equation describing such strong-field (ionization) processes. There are different techniques for solving the TDSE, including numerical, classical and semiclassical methods. One particularly intuitive method is the strong-field approximation (SFA), which provides insight into the ionization rates in strong laser fields \cite{faisal1973,reiss1980}. The SFA is a useful tool for studying the interactions between intense laser and atom as it allows for a simplified treatment of the interaction by assuming a classical description of the electromagnetic field, as opposed to utilizing a fully quantized treatment of the light-matter interaction. In comparison to other models \cite{torlina2012,javanainen1988,jheng2022,bauer2006,madsen2007,patchkovskii2016}, the SFA  is often favored for the calculation of angular and energy-resolved spectra. In this approximation, the Coulomb potential of the parent ion is disregarded in the electron continuum and only the electric field of the driving laser beam is considered. The transition amplitude can then be calculated by summing the direct and re-scattering amplitudes, with the assumption that the continuum wave function, including the laser field, is already known. This approximation has been utilized to determine both above-threshold ionization (ATI) and high-harmonic spectra for a range of laser beams in the near-infrared region \cite{milosevic2000,suarez2015,milosevic2016,Minneker,Paufler_2019}.

\begin{figure}[h]
	\includegraphics[width=0.48\textwidth]{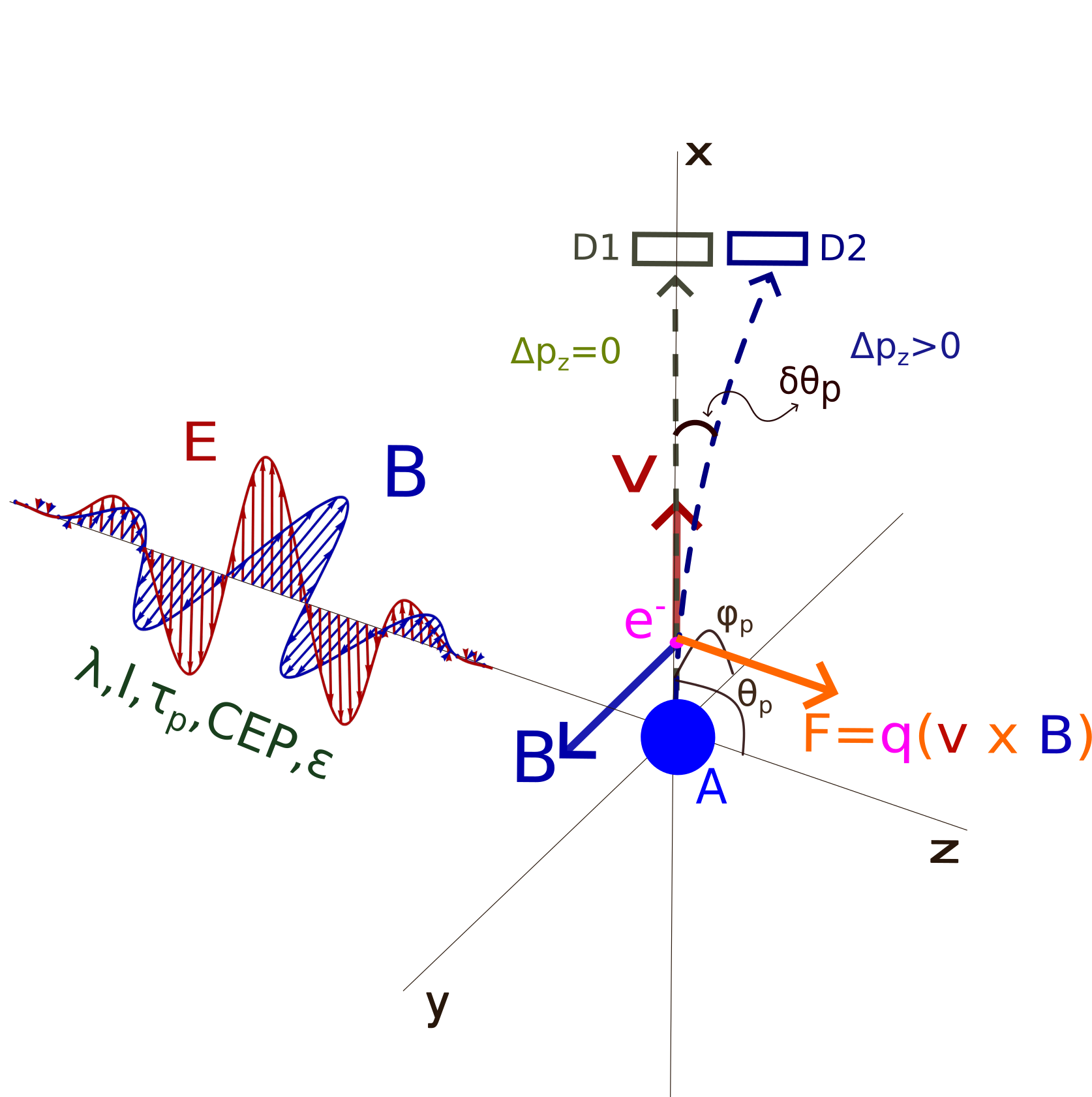}
	\caption{A typical setup for ATI measurement considered in our theory. A laser pulse with specified properties (wavelength $\lambda$, intensity I, pulse duration $\tau_{p}$, carrier envelope phase (CEP), and ellipticity ($\epsilon$) ionizes the atom A. The ionized electron with certain velocity ($v(t)=(v_{x},v_{y},0)$) is then accelerated by an electric field ($E(\boldsymbol{r},t)$) with asymptotic momentum ($\boldsymbol{p}=(p,\theta_{p},\varphi_{p})$) and experiences a Lorentz force ($F=q(v(t) \times B(\boldsymbol{r},t))$) due to a magnetic field ($B(\boldsymbol{r},t)$), leading to a shift in its momentum ($\Delta p_{z}$). This force causes the electron to be detected at a different location (D2) in the nondipole case, with a slight change in its polar angle ($\theta_{p}$) and a corresponding polar shift ($\pm\delta\theta_{p}$), assuming an azimuthal angle ($\varphi_{p}$) of 0.}
	\label{fig:lorentz}
\end{figure}

The maxima in the ATI spectra can typically be observed by placing a detector in the polarization plane perpendicular to the optical axis, as shown in Fig. (\ref{fig:lorentz}). In the dipole approximation, the magnetic part of the laser field is neglected, which means that the Lorentz force on the ionizing electron is not taken into account. Ignoring the Lorentz force is a valid assumption in low wavelength and low-intensity limit, where it has a negligible effect. However, in the case of long wavelength and more intense laser fields, the Lorentz force has a significant impact on the observed spectra and momentum distributions \cite{Fritzsche,Boning2}. Therefore, to measure the peak spectra accurately, the detector must be placed slightly away from the polar axis ($\theta_{\boldsymbol{p}}\pm \delta\theta_{p}$).

Recent measurements \cite{smeenk2011,ludwig2014,maurer2018,haram2019,hartung2019} of nondipole-induced peak shifts have sparked interest in the field. These experiments were conducted using femtosecond pulsed lasers. The focus of these experiments was on the overall shift of the momentum distribution along the laser's propagation direction.  The SFA has traditionally been used with laser fields in dipole approximation and has only recently been extended to nondipole scenarios \cite{boning2019,stefanie,Habibovi}.  However, such formalism is limited to plane-wave beams and the experiments mentioned above used lasers with durations of a few femtoseconds to measure shift in the momentum distribution. Therefore, a theory is needed that allows for more flexibility in the laser field and provides a better comparison with experimental measurements.

 This paper presents an extension of the previously introduced SFA (Ref. \cite{boning2019,rosenberg1993}) to incorporate not only the complex spatial structure but also the laser field envelope.  The paper is organized as follows: In Sec. ~\ref{sec:TF}, we discuss the theoretical formalism for the nondipole SFA, including an introduction to the SFA in Sec. ~\ref{subsec: SFA}. Followed by Sec. ~\ref{subsec: NDSFA} in which we discuss the continuum state derived for a monochromatic beam. The construction of the nondipole Volkov state with a vector potential containing both complex spatial structure and pulse envelope is discussed in Sec. ~\ref{subsec: MVS}. In Sec. ~\ref{subsec: TA}, we introduce the ionization amplitude that includes the formulated nondipole Volkov states. In Sec. ~\ref{sec: RnD}, we discuss our work by showing how the introduction of the pulse envelope can provide a better comparison with experiments. We also show the nondipole effects on the polar angular distribution of photoelectrons for different pulse durations. Finally, in Sec. ~\ref{sec:summary}, we summarize our work.\\
Note that atomic units ($m_{e}=\hbar=\frac{e^2}{4\pi\epsilon_{0}}=1$) are considered throughout the paper, unless stated otherwise.\\

\section{Theoretical Formalism}
\label{sec:TF}
\subsection{Strong-Field Approximation}
\label{subsec: SFA}

	The theoretical formalism for the nondipole description of the strong-field approximation is derived from the TDSE and describes the behavior of an electron in a strong electromagnetic field. The TDSE is derived using the Hamiltonian which includes the kinetic energy of photoelectron, classical representation of electromagnetic field that includes the effects of the field, including the magnetic part, on the electron's motion and the atomic binding potential. The solution of this equation describes the electron's wave function and its dynamics in the presence of a strong field. This semiclassical method provides a accurate description of the strong-field ionization processes and allows for the calculation of various physical observables, such as the ionization probability and the electron momentum distribution.
	To generalize this method, let us consider the transition of a bound electron described by state $\ket{\Psi_{0}(t)}$ evolved within a laser field at time t, with asymptotic momentum $\boldsymbol{p}$, into a continuum state $\ket{\Psi_{\boldsymbol{p}}(t)}$. Then the Schr\"o{}dinger equation can be written as
\begin{equation}
	\hat{H}\ket{\Psi(t)} = \dot{\iota} \frac{\partial}{\partial t}\ket{\Psi(t)},
	\label{eqn:1}
\end{equation}
with the total Hamiltonian given by
\begin{equation}
\hat{H} = \frac{\boldsymbol{\hat{p}}^{2}}{2} + \hat{V}_{le}(\boldsymbol{r},t) + \hat{V}(\boldsymbol{r}).
	\label{eqn:2}
\end{equation}
 Here, $\hat{V}(\boldsymbol{r})$ is the atomic binding potential and $\hat{V}_{le}(\boldsymbol{r},t)$ denotes the laser-electron interaction.\\
 The differential ionization probability for such an electron emitted with energy E$_{\boldsymbol{p}}=\frac{\boldsymbol{p}^{2}}{2}$ into the solid-angle d$\Omega_{\boldsymbol{p}}$ is given by
\begin{equation}
P(\boldsymbol{p})=\frac{|T_{\boldsymbol{p}}|^{2}d^{3}\boldsymbol{p}}{d\Omega_{\boldsymbol{p}}dE_{\boldsymbol{p}}}=p|T_{\boldsymbol{p}}|^{2},
	\label{eqn:3}
\end{equation}
where the bound-free transition amplitude $T_{\boldsymbol{p}}$, describing the transition of the electron from the ground state to a continuum, can be written as
\begin{equation}
T_{\boldsymbol{p}}=\lim_{t\to\infty,t'\to -\infty}\langle\Psi_{\boldsymbol{p}}(t)|\hat{U}(t,t')|\Psi_{0}(t')\rangle,
	\label{eqn:4}
\end{equation}
$\hat{U}(t,t')$  represents the time-evolution operator of the Hamiltonian (\ref{eqn:2}) that incorporates the  contributions of the binding potential and the laser field. To simplify above equation, we make use of the Dyson expansion. Thus we can write the transition amplitude (\ref{eqn:4}) as
\begin{widetext}
\begin{equation}
	\begin{aligned}
		T_{\boldsymbol{p}} =& (-\dot{\iota}) \lim_{t\to\infty,t'\to -\infty} \int_{t'}^{t}d\tau \langle\Psi_{\boldsymbol{p}}(t)|\hat{U}_{le}(t,\tau)\hat{V}_{le}(\boldsymbol{r},\tau)|\Psi_{0}(\tau)\rangle\\
		                    & +(-\dot{\iota})^2\lim_{t\to\infty,t'\to -\infty}\int_{t'}^{t}d\tau \int_{\tau}^{t}d\tau'\langle\Psi_{\boldsymbol{p}}(t)|\hat{U}(t,\tau')\hat{V}(\boldsymbol{r})\hat{U}_{le}(\tau',\tau)\hat{V}_{le}(\boldsymbol{r},\tau)|\Psi_{0}(\tau)\rangle.
	\end{aligned}
	\label{eq:5}
\end{equation}
\end{widetext}
To further simplify Eq. (\ref{eq:5}) we finally use the strong-field approximation (SFA) with the assumption as follows:\\
    1. The time evolution operator for the full Hamiltonian, $\hat{U}(t,t')$, approximately equals $\hat{U}_{le}(t,t')$ as the atomic potential is disregarded. The Hamiltonian $\hat{H}_{le} = \frac{\boldsymbol{\hat{p}}^{2}}{2} + \hat{V}_{le}(\boldsymbol{r},t) $ is associated with the time evolution operator $\hat{U}_{le}(t,t')$.\\
2. The final state $\ket{\Psi_{\boldsymbol{p}}(t)}$ is assumed as plane wave $\ket{\boldsymbol{p}}$ with $\braket{r|\boldsymbol{p}(t)} = (2\pi)^{-\frac{3}{2}}e^{\dot{\iota} \boldsymbol{p}\cdot r}e^{-\dot{\iota} E_{p}t}$\\
3. We consider initial state $\ket{\Psi_{0}(t)}$ to be an eigenstate of the atomic Hamiltonian $\hat{H}_0 = \frac{\boldsymbol{\hat{p}}^{2}}{2} + \hat{V}(\boldsymbol{r})$. This implies that the impact of the laser field on the bound state structure can be disregarded.\\
Considering the assumption (1), the time evolution of the electron-laser interaction Hamiltonian can be expanded in the complete basis of continuum states $\ket{\chi_{\boldsymbol{k}}(t)}$, characterized by the momenta $\boldsymbol{k}$,
\begin{equation}
\hat{U}_{le}(t,t')=\int d^{3}\boldsymbol{k} |\chi_{\boldsymbol{k}}(t)\rangle\ \langle \chi_{\boldsymbol{k}}(t')|.
	\label{eq:6}
\end{equation}
These assumptions enable us to write the transition amplitude as
		\begin{subequations}
			\begin{align}
				T_{\boldsymbol{p}} &= T^{(0)}_{\boldsymbol{p}}+T^{(1)}_{\boldsymbol{p}}\label{eq:7a},\\
				T^{(0)}_{\boldsymbol{p}} &= (-\dot{\iota}) \int_{-\infty}^{\infty}d\tau \langle \chi_{\boldsymbol{p}}(\tau)|\hat{V}_{le}(\boldsymbol{r},\tau)|\Psi_{0}(\tau)\rangle\label{eq:7b},\\
				\begin{split}
				T^{(1)}_{\boldsymbol{p}} &= (-\dot{\iota})^2\int_{-\infty}^{\infty}d\tau \int_{\tau}^{\infty}d\tau' \\ 
				&\langle \chi_{\boldsymbol{p}}(\tau')|\hat{V}(\boldsymbol{r})\hat{U}_{le}(\tau',\tau)\hat{V}_{le}(\boldsymbol{r},\tau)|\Psi_{0}(\tau)\rangle.
				\end{split}\label{eq:7c}
			\end{align}
		\label{eq:7}
		\end{subequations}
$T^{(0)}_{\boldsymbol{p}}$ and $T^{(1)}_{\boldsymbol{p}}$ refer to the direct and re-scattering transition amplitudes respectively.
\subsection{Nondipole continuum state}
\label{subsec: NDSFA}

 To account for the nondipole description of SFA to ATI, it is indispensable to write the vector potential of driving laser field with a $\boldsymbol{r}$-dependence. In other words, the vector potential varies with both space and time, giving rise to both an electric and a magnetic field. This complex dependence is necessary to accurately describe the behavior of the field and its effects on the electron dynamics in the strong-field approximation. For a laser field propagating with angular frequency $\omega$ in the direction of wave vector $\boldsymbol{k}=\frac{\omega}{c}\hat{k}$, the vector potential can be written as the superposition of plane-wave modes
\begin{equation}
	\begin{aligned}
		\boldsymbol{A}(\boldsymbol{r},t)&=\int d^{3}\boldsymbol{k}\boldsymbol{A}(\boldsymbol{k},t),\\
		\boldsymbol{A}(\boldsymbol{k},t)&=Re\{\boldsymbol{a}(\boldsymbol{k})e^{\dot{\iota} (\boldsymbol{k\cdot r}-\omega_{\boldsymbol{k}} t)}\}.
	\end{aligned}
	\label{eq:8}
\end{equation}
This form of vector potential, with $\boldsymbol{a}(\boldsymbol{k})$ as the complex Fourier coefficient, allows us to write the solution of the continuum state \cite{boning2019} as
\begin{equation}
	\chi_{\boldsymbol{p}}(\boldsymbol{r},t) = \frac{1}{(2\pi)^{\frac{3}{2}}}e^{-\dot{\iota}(E_{\boldsymbol{p}}t-\boldsymbol{p\cdot r})}e^{-\dot{\iota}\Gamma(\boldsymbol{r},t)},
	\label{eq:9}
\end{equation}
which is also called a modified Volkov state. It includes the particle's momentum $\boldsymbol{p}$ and the electromagnetic field, which is included in the modified Volkov phase $\Gamma(\boldsymbol{r},t)$, with the interaction between the two described by the classical Lorentz force equation. It is expressed as a superposition of plane wave solutions, each with a different momentum and energy. The Volkov state incorporates the effect of the electromagnetic field on the particle's motion through the phase factors of the plane waves and allows for the calculation of the particle's motion and energy in the presence of the electromagnetic field. The modified Volkov phase \cite{boning2019} is given by
\begin{widetext}
\begin{equation}
	\begin{aligned}
		\Gamma(\boldsymbol{r},t) = & \int d^{3}\boldsymbol{k} \rho_{\boldsymbol{k}}\sin(\text{u}_{\boldsymbol{k}}+\theta_{\boldsymbol{k}})\\
		& + \int d^{3}\boldsymbol{k} \int d^{3}\boldsymbol{k}'\left[ \alpha^{+}_{\boldsymbol{k},\boldsymbol{k}'}\sin(\text{u}_{\boldsymbol{k}}+\text{u}_{\boldsymbol{k}'}+\theta^{+}_{\boldsymbol{k},\boldsymbol{k}'}) + \alpha^{-}_{\boldsymbol{k},\boldsymbol{k}'}\sin(\text{u}_{\boldsymbol{k}}-\text{u}_{\boldsymbol{k}'}+\theta^{-}_{\boldsymbol{k},\boldsymbol{k}'})\right] \\
		& + \frac{1}{2}\int d^{3}\boldsymbol{k} \int d^{3}\boldsymbol{k}' \sigma_{\boldsymbol{k},\boldsymbol{k}'}\rho_{\boldsymbol{k}} \left(\frac{\sin(\text{u}_{\boldsymbol{k}}+\text{u}_{\boldsymbol{k}'}+\theta_{\boldsymbol{k}} +\xi_{\boldsymbol{k},\boldsymbol{k}'})}{\eta_{\boldsymbol{k}}+\eta_{\boldsymbol{k}'}} + \frac{\sin(\text{u}_{\boldsymbol{k}}-\text{u}_{\boldsymbol{k}'}+\theta_{\boldsymbol{k}} -\xi_{\boldsymbol{k},\boldsymbol{k}'})}{\eta_{\boldsymbol{k}}-\eta_{\boldsymbol{k}'}}\right).
	\end{aligned}
\label{eq:10}
\end{equation}
\end{widetext}
Here, $\rho_{\boldsymbol{k}},\theta_{\boldsymbol{k}},\sigma_{\boldsymbol{k},\boldsymbol{k}'}$ and $\xi_{\boldsymbol{k},\boldsymbol{k}'}$, used in Eq. (\ref{eq:10}) are the projection operators. These operators are dependent on the form of the Fourier coefficients, $\boldsymbol{a}(\boldsymbol{k})$, of the vector potential (\ref{eq:8}), as well as the photoelectron momentum ($\boldsymbol{p}$). The operators represent the projection of $\boldsymbol{p}$ and the wave vector $\boldsymbol{k}$ onto the $\boldsymbol{k}$-space vector potential $\boldsymbol{A}(\boldsymbol{k}',t)$. Additionally, the functions $\alpha^{\pm}_{\boldsymbol{k},\boldsymbol{k}'}$ represent the ponderomotive terms for each mode, which are based on the product of the Fourier coefficients, [$\boldsymbol{a}(\boldsymbol{k})\cdot \boldsymbol{a}(\boldsymbol{k}')$]. The definition of these operators can be found in Appendix \ref{Appendix B} and we also introduced $\eta_{\boldsymbol{k}}=\boldsymbol{p\cdot k}-\omega_{\boldsymbol{k}}$. 
\subsection{Modified Volkov states for a laser pulse}
\label{subsec: MVS}
Using laser pulses in above-threshold ionization experiments offers several benefits that make them an attractive choice for researchers. One of the key advantages is the high temporal and spatial resolution that can be achieved using laser pulses. With duration on the order of femtoseconds or even attoseconds, laser pulses can provide precise time-resolved measurements of the ionization process.
To derive the continuum wave function for a plane wave pulse propagating in $\hat{z}$-direction, we start with the vector potential given by
\begin{equation}
	\begin{aligned}
		\boldsymbol{A}(\boldsymbol{r},t) = \frac{A_{0}}{\sqrt{1+\epsilon^{2}}}f(\boldsymbol{r},t)
		\biggl(&\cos(\text{u} + \phi_{\text{cep}})\boldsymbol{e}_{x}  \\ + \epsilon\Lambda
		&\sin(\text{u} + \phi_{\text{cep}})\boldsymbol{e}_{y}\biggr)
		\label{eq:11}
	\end{aligned}
\end{equation}
Here, we introduce the variable $u=\boldsymbol{k}_{0}\cdot\boldsymbol{r}-\omega_{0}t$, representing the phase of the vector potential. The envelope function $f(\boldsymbol{r},t)$ describes the shape of the pulse. $\epsilon\text{,}\quad\Lambda\text{,}\quad\phi_{\text{cep}}$ are the corresponding ellipticity, helicity and carrier  envelope phase respectively. Here, we will focus on the sin-squared envelope, given by:
\begin{equation}
f(\boldsymbol{r},t) = \begin{cases}
	\sin^{2}(\frac{\text{u}}{2\text{n}_{\text{p}}}), & 0\leq t \leq \tau_{p}\\
		0,                                & \mathrm{otherwise.}
	\end{cases}
\label{eq:12}
\end{equation}
The parameter $\text{n}_{\text{p}}$ represents the number of optical cycles contained within the pulse duration of $\tau_{p}$. Now substituting Eq. (\ref{eq:12}) into Eq. (\ref{eq:11}) and expanding the trigonometric products, we obtain the vector potential:
\begin{equation}
	\begin{aligned}
		\boldsymbol{A}(\boldsymbol{r},t) = \sum_{j=-1}^{1} \frac{A_{j}}{\sqrt{1+\epsilon^{2}}}
		\biggl(&\cos(\text{u}_{j}+\phi_{\text{cep}})\boldsymbol{e}_{x} \\
		+\epsilon\Lambda &\sin(\text{u}_{j} + \phi_{\text{cep}})\boldsymbol{e}_{y}\biggr)
	\end{aligned}
	\label{eq:13}
\end{equation}
The vector potential is expressed as the superposition of of three monochromatic plane-wave beams with different frequencies that are labeled by $j$. For example, when $j=-1$ then $\omega=(1-1/\text{n}_{\text{p}})\omega_{0}$. Similarly, for $j=0 $ and $ 1$, $\omega = \omega_{0}$ and $\omega=(1+1/\text{n}_{\text{p}})\omega_{0}$ respectively.  The  quantities with the indices $j's$ can be found in Appendix \ref{Appendix A}. The decomposition allows us to write the vector potential in the form of Eq. (\ref{eq:8}). Now, we can solve Eq. (\ref{eq:10}) using vector potential (\ref{eq:13})  and write the modified Volkov phase in the form
\begin{widetext}
\begin{equation}
	\begin{aligned}
		\Gamma(r,t) =& -\rho_{\epsilon}\sum_{j=-1}^{1}\frac{D_{j}}{C_{j}}\sin(C_{j}\text{u} + \phi_{\text{cep}}-\Lambda\varphi^{(\epsilon)}_{p}) -\alpha \frac{1-\epsilon^{2}}{1+\epsilon^{2}}\sum_{j=-1}^{1}\frac{D^{2}_{j}}{2C_{j}}\sin(2C_{j}\text{u}+2\phi_{\text{cep}})\\
		& -2\alpha\sum\limits_{(i,j)=(0,-1),(0,1),(-1,1)}\left[ \frac{1-\epsilon^{2}}{1+\epsilon^{2}}\frac{D_{i}D_{j}}{C_{i}+C_{j}}\sin\Bigl((C_{i}+C_{j})\text{u}+2\phi_{\text{cep}}\Bigr) -\frac{D_{i}D_{j}}{C_{i}-C_{j}}\sin\Bigl((C_{i}-C_{j})\text{u}\Bigr)\right]\\
		& -\alpha \sum_{j=-1}^{1} D^{2}_{j} u
	\end{aligned}.
	\label{eq:14}
\end{equation}
\end{widetext}
Here we define $$\alpha = \frac{U_{p}}{\omega_{0}}\frac{1}{1-\boldsymbol{p}\cdot\boldsymbol{k}_{0}/\omega_{0}},$$ as the modified pondermotive energy and $$\rho_{\epsilon}=\frac{A_{0}}{\sqrt{1+\epsilon^{2}}}\frac{\rho^{(\epsilon)}}{\omega_{0}(1-\boldsymbol{p}\cdot\boldsymbol{k}_{0}/\omega_{0})},$$ as the product of kinetic and field-induced photoelectron momentum. The constants $D_{j}$ and $C_{j}$ are defined in Appendix \ref{Appendix A}. The general solution of Eqs.~\eqref{eq:10} to \eqref{eq:14} can be found in Appendix \ref{Appendix B}. $\boldsymbol{p}^{(\epsilon)}$ is a vector in Cartesian coordinates, represented as $(p_{x},\quad\epsilon p_{y},\quad 0)$, which describes the modulus and azimuthal angle of an auxiliary and polarization-dependent momentum vector. The vector can also be represented in spherical coordinates as $(p^{(\epsilon)},\quad \pi/2,\quad \varphi_{p}^{(\epsilon)})$. The vector lies within the polarization plane. The summation over $j$ refers to the situation where we observe the superposition of plane-wave beams resulting from the vector potential (\ref{eq:13}).  For a plane wave pulse, the solution can be represented as a combination of 3-color monochromatic plane wave beam solutions by insert equation (\ref{eq:14}) into equation (\ref{eq:9}). The full solution, known as the Volkov state, can then be written as a superposition of individual terms, corresponding to each "j", given by
\begin{widetext}
\begin{equation}
	\chi_{\boldsymbol{p}}(\boldsymbol{r},t) = (2\pi)^{-3/2}\prod_{\substack{j=-1}}^{1}\left[ \prod_{\substack{i=1}}^{4}\sum_{n_{i}^{j}=-\infty}^{\infty} J_{n_{i}^{j}}(x_{i}^{j}) \times e^{-\dot{\iota} (E_{N}t- \boldsymbol{p}_{N}\cdot\boldsymbol{r}-{\Phi_{N}})}\right].
	\label{eq:155}
\end{equation}
\end{widetext}

Here, we used the Jacobi-Anger expansion to decompose the Volkov phase (\ref{eq:9}). In the above equations, the integer $n_{i}^{j}$ defines the number of photons absorbed as a result of the superposition of three-color beams. The value of argument ($x_{i}^{j}$), for different values of i and j, of the  Bessel function are given in Table \ref{tab:table1}.
{\makeatletter
	\renewcommand\table@hook{\large}
	\makeatother
\begin{table}
	\caption{Arguments ($x_{i}^{j}$) of the Bessel functions in Eq. (\ref{eq:21} and \ref{eq:155}). The indices i and j are counted in the columns and rows, respectively.}
		\label{tab:table1}
		\begin{ruledtabular}
		\begin{tabular}{c|cccc} 
			 $x_{i}^{j}$&1&2 &3&4\\ \hline \\[0.01cm]
			 -1 &$\rho_{\epsilon}\frac{D_{-1}}{C_{-1}}$ & $\alpha\frac{1-\epsilon^{2}}{1+\epsilon^{2}}\frac{D^{2}_{-1}}{2C_{-1}}$ &$2\alpha\frac{1-\epsilon^{2}}{1+\epsilon^{2}}\frac{D_{0}D_{1}}{C_{0}+C_{1}}$  &$2\alpha\frac{D_{0}D_{1}}{C_{0}-C_{1}}$\\[0.5cm] 
			  0 & $\rho_{\epsilon}\frac{D_{0}}{C_{0}}$ &$\alpha\frac{1-\epsilon^{2}}{1+\epsilon^{2}}\frac{D^{2}_{0}}{2C_{0}}$ &$2\alpha\frac{1-\epsilon^{2}}{1+\epsilon^{2}}\frac{D_{0}D_{-1}}{C_{0}+C_{-1}}$ &$2\alpha\frac{D_{0}D_{-1}}{C_{0}-C_{-1}}$\\[0.5cm]
			 1 &$\rho_{\epsilon}\frac{D_{1}}{C_{1}}$  & $\alpha\frac{1-\epsilon^{2}}{1+\epsilon^{2}}\frac{D^{2}_{1}}{2C_{1}}$ &$2\alpha\frac{1-\epsilon^{2}}{1+\epsilon^{2}}\frac{D_{-1}D_{1}}{C_{-1}+C_{1}}$ &$2\alpha\frac{D_{-1}D_{1}}{C_{-1}-C_{1}}$\\[0.5cm]
		\end{tabular}
	\end{ruledtabular}
\end{table}}
Also, the modified photoelectron energy ($E_{N}$), momentum ($\boldsymbol{p}_{N}$) and phase($\Phi_{N}$) are given as\\
\begin{subequations}
	\begin{align}
		E_{N} &= E_{\boldsymbol{p}} + N^{(0)}\omega_{0},\\
		\boldsymbol{p}_{N} &= \boldsymbol{p} + N^{(0)} \boldsymbol{k}_{0},\\
		\Phi_{N} &= (N^{(1)}+2N^{(2)})\phi_{\text{cep}} +N^{(1)}\Lambda\varphi^{(\epsilon)}_{p},
	\end{align}
\end{subequations}
respectively.  Here, we introduced short notation $N^{(0,1,2)}$, which depends on the pulse cycle and the $n^j_i$, that occur in Bessel functions and are given as 
	\begin{equation}
		\begin{aligned}
			N^{(0)}= \frac{3\alpha}{8} &+ (n^{1}_{0}+2n^{2}_{0}+n^{3}_{0}+n^{3}_{-1}+n^{4}_{0}+n^4_{-1})C_{0}\\
			&+ (n^1_{-1}+2n^2_{-1}+n^3_{0}+n^3_{1}-n^4_{0}+n^4_{1})C_{-1}\\
			&+ (n^1_{1}+2n^2_{1}+n^3_{-1}+n^3_{1}-n^4_{-1}-n^4_{1})C_{1},\\
			N^{(1)}=n^1_{0}&+n^1_{-1}+ n^1_{1},\\
			N^{(2)}=n^2_{0}&+n^2_{-1}+n^2_{1}+n^3_{0}+n^3_{-1}+n^3_{1}.
		\end{aligned}
	\end{equation}
\subsection{Transition Amplitude}
\label{subsec: TA}
The direct transition amplitude determines the probability that an electron undergo a transition from a bound initial state to a continuum states when interacting a strong electric field. We can solve for this probability using the equation (\ref{eq:7b}). As the operator $\hat{V}_{le}= \hat{H}_{le}-\hat{H}_{A}+V(\boldsymbol{r})$, we can further simplify the direct amplitude by performing the integration by parts. Since the vector potential is nonzero within some interval $t_{i}\leq t\leq t_{f}$, together with the operator $\hat{V}_{le}$, the original equation (\ref{eq:7b}) is modified to a more simplified form as
\begin{equation}
	\begin{aligned}
		T^{(0)}_{\boldsymbol{p}}=&-\dot{\iota} \int_{t_{i}}^{t_{f}}d\tau \left( \langle \chi_{\boldsymbol{p}}(\tau)|\frac{\overleftarrow{\partial}}{\partial\tau}+\frac{\overrightarrow{\partial}}{\partial\tau}|\Psi_{0}(\tau)\rangle \right)\\
		&-\dot{\iota}\int_{t_{i}}^{t_{f}}d\tau \langle \chi_{\boldsymbol{p}}(\tau)|V(\boldsymbol{r})|\Psi_{0}(\tau)\rangle \\
		=&-\langle \chi_{\boldsymbol{p}}(\tau)|\Psi_{0}(\tau)\rangle|^{t_{f}}_{t_{i}}-\dot{\iota}\int_{t_{i}}^{t_{f}}d\tau \langle \chi_{\boldsymbol{p}}(\tau)|V(\boldsymbol{r})|\Psi_{0}(\tau)\rangle.
	\end{aligned}
	\label{Eq:19}
\end{equation}
 For a finite laser pulse, the time period during which the pulse exists is determined by the pulse duration. In the previous case, as stated in Eq. (\ref{eq:12}), the pulse begins at time $t_{i} = 0$ and ends at time $t_{f} = \tau_{p}$. The interval of the pulse can be thought of as the duration of the pulse, which is specified by the difference between the starting and ending times. The pulse duration defines the time period during which the pulse is present. For a hydrogen-like 1s wave-function we can easily replace the initial bound state with a modified ionization potential $I_{p}$ in a way that
\begin{equation}
	\ket{\Psi_{0}(t)}=\ket{\Psi_{0}}e^{\dot{\iota}I_{p}t}=\frac{2I_{p}^{\frac{3}{2}}}{\sqrt{\pi}}e^{-\sqrt{2I_{p}}\boldsymbol{r}}e^{\dot{\iota} I_{p}t}.
	\label{eq:20}
\end{equation}
Upon insertion of this expression together with the Volkov state (\ref{eq:155}) into Eq. (\ref{Eq:19}), we finally arrive at our final expression for the transition amplitude:
\begin{widetext}
	\begin{equation}
		\begin{aligned}
			T^{(0)}_{\boldsymbol{p}}=& \prod_{\substack{j=-1}}^{1}\left[\prod_{\substack{i=1}}^{4}\sum_{n_{i}^{j}=-\infty}^{\infty} J_{n_{i}^{j}}(x_{i}^{j})\times\left[\frac{\langle \boldsymbol{p}_{N}|V(\boldsymbol{r})|\Psi_{0}\rangle}{I_{p}+E_{N}} + \langle \boldsymbol{p}_{N}|\Psi_{0}\rangle\right]\times \left[1-e^{\dot{\iota}(I_{p}+E_{N})\tau_{p}}\right]\right],
		\end{aligned}
	\label{eq:21}
	\end{equation}

where the matrix element of the Coulomb potential and the momentum-space initial wave function are given by
\begin{equation}
	\begin{aligned}
		\langle \boldsymbol{p}_{N}|V(\boldsymbol{r})|\Psi_{0}\rangle&=-\frac{2^{\frac{3}{2}}I_{p}^{\frac{5}{4}}}{\pi}\frac{1}{\frac{\boldsymbol{p}_{N}^{2}}{2}+ I_{p}},\\
		\langle \boldsymbol{p}_{N}|\Psi_{0}\rangle&=\frac{2I_{p}^{\frac{5}{4}}}{\pi\sqrt{2}}\frac{1}{\left(\frac{\boldsymbol{p}_{N}^{2}}{2}+I_{p}\right)^{2}},
	\end{aligned}
\end{equation}
respectively. The momentum-space initial wave function is solved by taking the Fourier transform of initial state wave-function (\ref{eq:20}).
\end{widetext}
\onecolumngrid\
\begin{figure*}	    
	\centering\includegraphics[width=17cm]{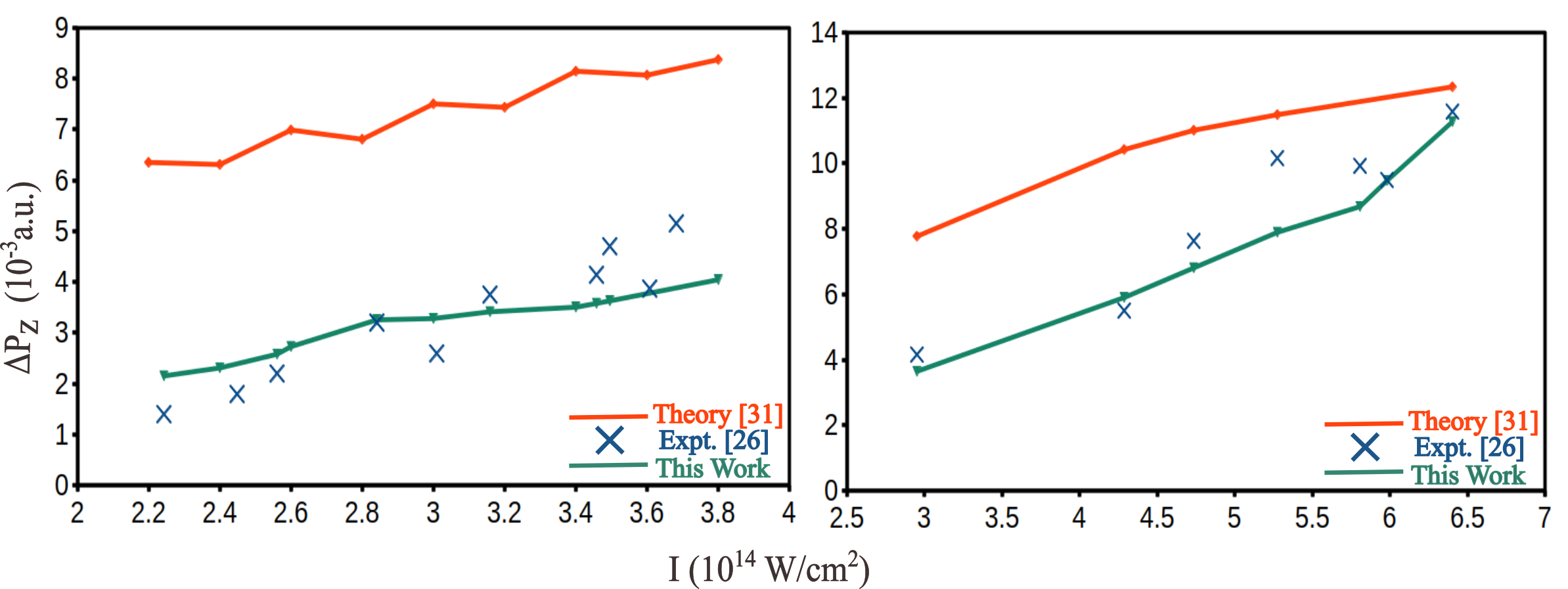}
	\caption{The peak shift $\Delta P_{z}$ of the maxima in ATI spectra are plotted as a function of laser intensity I for a circularly polarized 800 nm, 15 fs laser pulse. Results are shown for two different atomic targets, Ar (left) and Ne (right), and are compared to previous experimental (blue-crosses) and theoretical work: orange (Ref. \cite{boning2019}), blue (Ref. \cite{smeenk2011}), and green (this work).}
	\label{fig:2}
\end{figure*}
\begin{figure*}
	\centering\includegraphics[width=17cm]{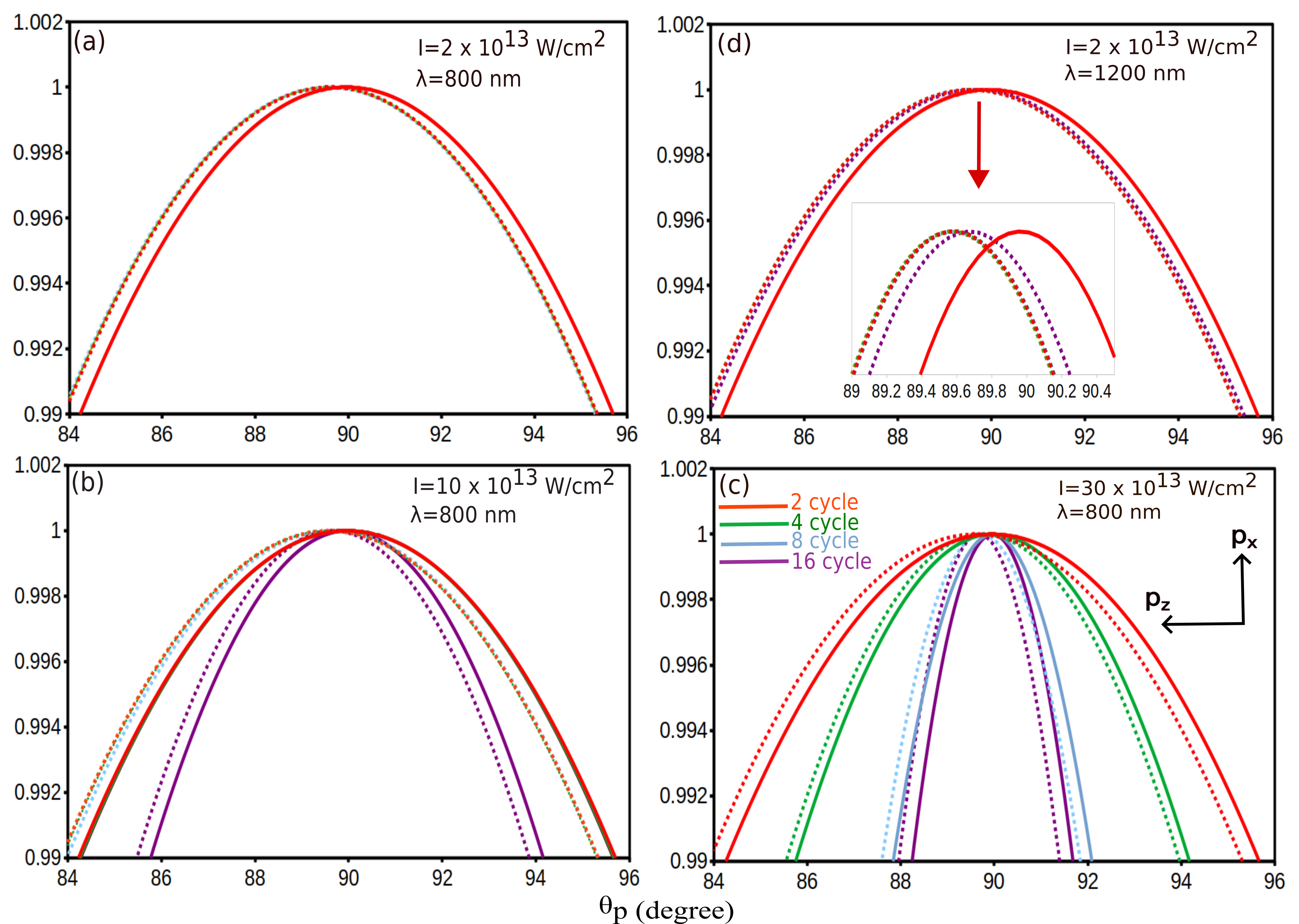}
	\caption{Normalized polar angular distribution (PAD) in the propagation ($p_{x}-p_{z}$) plane ($\varphi_p= 0$) for argon interacting with a circularly polarized sine-squared pulse. The laser wavelengths of 800 nm and 1200 nm are used with intensities ranging between $I=2\times10^{13} - 30\times10^{13} \text{W/cm}^{2}$. The pulse duration is varied with different numbers of optical cycles ($\text{n}_{\text{p}}=2, 4, 8, 16$) and photoelectron energy ($E_{\text{p,max}}$), at which the maximum ionization probability occurs, is kept constant. The Lorentz force acts on the electron, causing it to be pushed in the direction of laser propagation. The results from both dipole (solid curve) and nondipole (dotted curve) computations are shown.}
	\label{fig:3}
\end{figure*}	

\twocolumngrid\

\section{Results and Discussion}
\label{sec: RnD}

The differential ionization probability (\ref{eqn:3}) can be computed by utilizing the transition amplitude (\ref{eq:21}) derived in the previous section. If we consider the ionization probability as a function of photoelectron energy, with fixed laser parameters, we can compute the ATI spectrum. The above formalism supports both linearly and elliptically polarized laser fields. In case of a circularly polarized ($\epsilon=1$) laser field, the Bessel functions with indices $n_{i}^{j}$, $j=(-1,0,1)$ and $i=(2,3)$ can be ignored, using the fact that $J_{n}(0)=\delta_{n,0}$, reducing the Eq. (\ref{eq:21}) with six Bessel functions containing terms $n_{i}^{j}$, $j=(-1,0,1)$ and $i=(1,4)$. In order to further simplify the infinite sum involving Bessel functions, we exploit the properties of these functions i.e., for $n_{i}^{j}>x_{i}^{j}$, $J_{n_{i}^{j}}(x_{i}^{j})\sim e^{-n_{i}^{j}}$. As a result, we can establish a cut-off of our Bessel function as the greatest integer that is slightly greater than the argument of the function i.e, $n_{max}=\lceil x_{i}^{j}\rceil$.

The use of a circularly polarized light has several benefits over a linearly polarized field.
First, a circularly polarized field prevents Coulomb focusing \cite{comtois1923observation}. In the Coulomb focusing, the electrons are attracted towards the charge center of the atom due to the Coulomb force, which result in a narrowing or focusing of the electron momentum distribution. So, the Coulomb focusing can be a problem in experiments that aim to study the momentum distribution of emitted electrons, as it can obscure the underlying physics. By using circularly polarized light, we can avoid Coulomb focusing and obtain a more accurate representation of the electron momentum distribution.
Second, circularly polarized light provides a simple way to measure the laser intensity. In other words, for a circular polarized light, the electron momentum distribution takes the form of a torus \cite{Smeenk_2009}, where the radius of distribution is proportional to the peak electric field experienced by the neutral atoms. This property of circularly polarized light can be useful for understanding the dynamics of the ATI process and for characterizing the laser intensity.

The peak shift ($\Delta P_{z}$) in ATI is determined by the intensity and frequency of the incident laser pulse, as well as the properties of the target atom. In general, the peak offset increases with increasing laser intensity because higher intensities lead to stronger interactions between the laser field and the target electrons. The peak offset is an important parameter in the study of ATI, as it provides insight into the dynamics of the ionization process and the underlying physical mechanisms. As illustrated in Fig. \ref{fig:2}, a comparison of peak shifts for two different cases, Argon and Neon, is presented alongside experimental and theoretical results for a monochromatic plane-wave beam. While the number of points depicted in the plots is limited, the extended computation time facilitated a more thorough and accurate analysis. Fig. \ref{fig:2} clearly shows that for a plane-wave beam, the peak shift largely differs from the experimental findings while our theoretical results for a pulse show a reduced discrepancy.

The parameters of laser pulse, including pulse duration \cite{Freeman}, intensity \cite{Lompre}, and wavelength \cite{Marchenko_2010}, can alter the ATI spectra. These changes in the spectra can also affect the polar angular distribution of the emitted electrons. For a single-cycle pulse, the angular distribution of the photoelectrons is isotropic, i.e., the electrons are ejected equally in all directions. In other words, by considering that the laser field is strong enough to ionize the atom in a single cycle, and the electron does not have time to interact with the laser field for a prolonged period, the electron is liberated from the atom at a random point in the laser cycle, leading to an isotropic polar angular distribution. In contrast to single cycle pulse, when the pulse duration is longer than one cycle, the polar angular distribution of the photoelectrons can be anisotropic as seen in Fig. \ref{fig:3}. For a long pulse, the electron has time to interact with the laser field for multiple cycles, and its motion is affected by the shape of field. The characteristics of a pulse result in the photoelectrons being preferentially emitted in certain directions, which include both the pulse duration and the laser field strength. When it comes to the effect of wavelength, short wavelength laser light leads to a more anisotropic polar angular distribution of the emitted electrons, while long wavelength laser light leads to a more isotropic distribution. Such effect in the polar distribution arises because, in a short wavelength, a laser pulse is more tightly focused in time, making it more likely to cause electrons to be emitted in a specific direction. Long wavelength laser light, on the other hand, is less tightly focused and tends to produce a more isotropic distribution. Regarding to intensity \cite{Feldmann1987}, as the intensity of photons increases, the electrons are affected by the laser's electric field for a long period, causing them to be preferentially emitted in certain directions, closely following the direction of the electric vector of the radiation field. The number of pulse cycles, wavelength, and intensity of the laser light all interact in a complex way to determine the polar angular distribution of ATI electrons.

In Fig. \ref{fig:3}, we present a detailed analysis of the effect of laser pulse intensity, wavelength, and duration on the polar angular distribution in the propagation plain of photoelectrons emitted in the process of ATI. Our results indicate that the pulse duration plays a crucial role in determining the angular distribution of photoelectrons, particularly when the laser intensity is high and the wavelength is short. In Figure  \ref{fig:3}(a), we observe that for different pulse cycles, the results of both dipole and non-dipole computations overlap, making it challenging to differentiate the results. In other words, at low intensities and wavelengths, the pulse cycles has no effect on the polar angular distribution.  However, when the intensity of the laser pulse is increased and keeping the wavelength low, as shown in Figure \ref{fig:3}(b-c), the results become increasingly distinct for different pulse cycles. The polar angular distribution in the plane of propagation is taken by keeping the photo-electron energy constant at the maximum. Since we change the pulse duration the peak maximum of energy of the photo electron slights change because ATI peaks breaks up into a series of extremely narrow lines \cite{Freeman}. At longer wavelengths, the change is minimum and increases with shorter wavelengths and higher intensities. In contrast to intensity, Figure \ref{fig:3}(d) illustrates that such effects do not occur when the wavelength is increased. Furthermore, the non-dipole effects remain relatively unchanged with varying pulse durations. For the longest pulses considered with long wavelengths or high intensities, the magnetic component loses its significance. These findings indicate that in monochromatic fields, the ionization yield is not affected by the magnetic field, and the dipole and non-dipole results tend to converge, as previously established in the literature \cite{Simonsen2015}.


\section{Summary}\label{sec:summary}
\label{sec: sum}
The nondipole strong-field approximation (NSFA) is a method for describing the behavior of electrons in intense laser fields. The classical representation of a laser field is considered, and the interaction of the electron with the field is described using perturbation theory. This approach allows for the calculation of the momentum and kinetic energy of photoelectrons, which can be measured experimentally. The photoelectron momentum shift is a phenomenon observed in experiments using few-cycle pulse, where the momentum of the photoelectron is altered towards a positive direction in relation to the propagation of the laser pulse due to its magnetic component.

Our results demonstrate that the NSFA with extension to few cycle limit we proposed is a valuable theoretical method for calculating the effects of strong electromagnetic fields on the behavior of electrons. By assuming that the field is a few-cycle pulse, the NSFA allows for the calculation of the photoelectron momentum shift correctly, which is an important measure of the field's strength. We showed how the inclusion of a finite pulse duration in such a theory can significantly enhance the prediction of nondipole effects. The NSFA provides a simple and accurate way to calculate the peak shift, and is therefore crucial for understanding the behavior of electrons in strong-fields. The parameters of the laser pulse, including pulse duration, intensity, and wavelength, can alter the ATI spectra and affect the polar angular distribution of the emitted electrons. The results of our analysis provide valuable insights into the underlying physics of ATI and can inform the design of future experiments aimed at studying this phenomenon.


\begin{acknowledgments}
This work has been funded by the Deutsche Forschungsgemeinschaft (DFG, German Research Foundation)—440556973 and also by the Research School of Advanced Photon Science (RS-APS) of Helmholtz Institute Jena, Germany. We would like to express our sincere gratitude to Dr. Birger B\"o{}ning for his invaluable contribution to the direction of this research.

\end{acknowledgments}

\appendix 

\section[Appendix A]{Vector Potential}\label{Appendix A}
It is convenient to represent the electromagnetic field, particularly when working with time-varying fields or in situations where the electric and magnetic fields are orthogonal, in terms of a vector field.
In case of a sine squared pulse, the vector potential is given by the following equation:

\begin{equation}
	\begin{aligned}
		\boldsymbol{A}(\boldsymbol{r},t) = \frac{A_{0}}{\sqrt{1+\epsilon^{2}}}f(\boldsymbol{r},t)\biggl(
		&\cos(\boldsymbol{k}_{0}\cdot\boldsymbol{r}-\omega_{0}t + \phi_{\text{cep}})\boldsymbol{e_{x}} \\
		+ \epsilon\Lambda &\sin(\boldsymbol{k}_{0}\cdot\boldsymbol{r}-\omega_{0}t + \phi_{\text{cep}})\boldsymbol{e_{y}}\biggr).
	\end{aligned}
\label{Eq:A1}
\end{equation}
$f(\boldsymbol{r},t)$ is an envelope function that varies sinusoidally over time and is given by  
\begin{equation}
	f(\boldsymbol{r},t) = \begin{cases}
		\sin^{2}(\frac{\boldsymbol{k}_{0}\cdot\boldsymbol{r}-\omega_{0}t}{2\text{n}_{\text{p}}}), & 0\leq t \leq \tau_{p}\\
		0,                                & \mathrm{otherwise}.
	\end{cases}
\label{Eq:A2}
\end{equation}
Now if we expand the trigonometric products, inserting (\ref{Eq:A2}) into (\ref{Eq:A1}) , the vector potential can be written as
\begin{widetext}
\begin{equation}
	\begin{aligned}
		\boldsymbol{A}(\boldsymbol{r},t)=\frac{A_{0}}{\sqrt{1+\epsilon^{2}}}\Bigg[-\frac{1}{4}&\Bigg(\cos\Big(\text{u}(1-\frac{1}{\text{n}_{\text{p}}})+\phi_{\text{cep}}\Big) + \cos\Big(\text{u}(1+\frac{1}{\text{n}_{\text{p}}})+\phi_{\text{cep}}\Big) - 2\text{cos}(\text{u}+\phi_{\text{cep}}) \Bigg)\boldsymbol{e}_{x}\\
		 -\frac{\epsilon\Lambda}{4}&\Bigg(\sin\Big(\text{u}(1-\frac{1}{\text{n}_{\text{p}}})+\phi_{\text{cep}}\Big)+ \sin\Big(\text{u}(1+\frac{1}{\text{n}_{\text{p}}})+\phi_{\text{cep}}\Big) -2\text{sin}(\text{u}+\phi_{\text{cep}}) \Bigg)\boldsymbol{e}_{y} \Bigg].
	\end{aligned}
    \label{Eq:A3}
\end{equation}
\end{widetext}
In more compact form, we can write
\begin{equation}
	\begin{aligned}
		\boldsymbol{A}(\boldsymbol{r},t) = \sum_{j=-1}^{1} \frac{A_{j}}{\sqrt{1+\epsilon^{2}}} \biggl(
		&\cos(\text{u}_{j}+\phi_{\text{cep}})\boldsymbol{e}_{x}  \\
		+ \epsilon\Lambda &\sin(\text{u}_{j} + \phi_{\text{cep}})\boldsymbol{e}_{y}\biggr),
	\end{aligned}
   \label{Eq:A4}
\end{equation}
where we used the short notation 
\begin{equation}
	\begin{aligned}
		\text{u}_{j}&=\boldsymbol{k}_{j}\cdot\boldsymbol{r}-\omega_{j}t=C_{j}(\boldsymbol{k}_{0}\cdot\boldsymbol{r}-\omega_{0}t),\\
		A_{j}&=D_{j}A_{0},
	\end{aligned}
\end{equation}
and $j$ ranging from -1 to +1. Specifically, the constants C's and D's are given by
\begin{equation}
	\begin{aligned}
		C_{0}&=1,\quad C_{-1}=1-\frac{1}{\text{n}_{\text{p}}}, \quad C_{1}=1+\frac{1}{\text{n}_{\text{p}}},\\
		D_{0}&=\frac{1}{2} \quad \text{and} \quad D_{-1}=D_{1}=\frac{-D_{0}}{2}.
	\end{aligned}
\end{equation}
\section[Appendix B]{Modified Volkov Phase}\label{Appendix B}
In this appendix, we provide a more thorough treatment of the derivation of the nondipole Volkov states. This was briefly discussed in Section \ref{subsec: NDSFA}. Specifically, we will show how to solve for the vector potential, which is typically written in terms of plane waves, for each value of $j$ [of Eq. (\ref{Eq:A4})]. For a vector potential of type (\ref{eq:8}), the Volkov solution of an outgoing photoelectron that takes into account nondipole effects is given by \cite{boning2019} 
\begin{equation}
	\chi_{\boldsymbol{p}}(\boldsymbol{r},t) = \frac{1}{(2\pi)^{\frac{3}{2}}}e^{-\dot{\iota}(E_{p}t-\boldsymbol{pr})}e^{-\dot{\iota}\Gamma(\boldsymbol{r},t)},
\end{equation}
which includes a modified Volkov phase term (\ref{eq:10}). To determine the individual functions $\lambda_{\boldsymbol{k}},\theta_{\boldsymbol{k}} ,  \rho_{\boldsymbol{k}} , \alpha^{\pm}_{\boldsymbol{k},\boldsymbol{k}'}, \theta^{\pm}_{\boldsymbol{k},\boldsymbol{k}'}, \xi_{\boldsymbol{k},\boldsymbol{k}'}, $ and $ \sigma_{\boldsymbol{k},\boldsymbol{k}'}$ in the Volkov phase for a given kinetic momentum $\boldsymbol{p}$ of the photoelectron, it is most efficient to use the plane-wave amplitudes $\boldsymbol{A}(\boldsymbol{k},t)$ and $ \boldsymbol{a}(\boldsymbol{k})$ and evaluate the left-hand sides of the following expressions \cite{boning2019}
\begin{widetext}
\begin{equation}
	\begin{aligned}
		\boldsymbol{p}\cdot\boldsymbol{A}(\boldsymbol{k},t)&=\lambda_{\boldsymbol{k}}\text{cos}(\text{u}_{\boldsymbol{k}}+\theta_{\boldsymbol{k}}),\qquad
		-\boldsymbol{k}\cdot A(\boldsymbol{k}',t)=\sigma_{\boldsymbol{k},\boldsymbol{k}'}\text{cos}(\text{u}_{\boldsymbol{k}'}+\xi_{\boldsymbol{k},\boldsymbol{k}'}),\\
		\frac{1}{4}\boldsymbol{a}(\boldsymbol{k})\cdot \boldsymbol{a}(\boldsymbol{k}')&=\Delta^{+}_{\boldsymbol{k},\boldsymbol{k}'}\text{exp}(\dot{\iota}\theta^{+}_{\boldsymbol{k},\boldsymbol{k}'}),\qquad
		\frac{1}{4}\boldsymbol{a}(\boldsymbol{k})\cdot \boldsymbol{a}^*(\boldsymbol{k}')=\Delta^{-}_{\boldsymbol{k},\boldsymbol{k}'}\text{exp}(\dot{\iota}\theta^{-}_{\boldsymbol{k},\boldsymbol{k}'}),\\
		\rho_{\boldsymbol{k}}=\frac{\lambda_{\boldsymbol{k}}}{\eta_{\boldsymbol{k}}} &\quad\text{and}\quad
		\alpha^{\pm}_{\boldsymbol{k},\boldsymbol{k}'}=\frac{\Delta^{\pm}_{\boldsymbol{k},\boldsymbol{k}'}}{\eta_{\boldsymbol{k}}\pm\eta_{\boldsymbol{k}'}}.
		\label{Eq:B3}
	\end{aligned}
\end{equation}
These equations reveal that $\rho_{\boldsymbol{k}}$ represents the product of the kinetic and field-induced momenta $\boldsymbol{p}\cdot\boldsymbol{A}(\boldsymbol{k},t)$ for a specific Fourier mode $\boldsymbol{A}(\boldsymbol{k},t)$. Additionally, the functions $\alpha^{\pm}_{\boldsymbol{k},\boldsymbol{k}'}$ represent the ponderomotive terms for each mode, given by $\boldsymbol{a}(\boldsymbol{k})\cdot \boldsymbol{a}(\boldsymbol{k}')$. For the vector potential (\ref{Eq:A4}), the above Eqs. (\ref{Eq:B3}) can be solved for individual $j$ values. Thus giving
\begin{equation}
	\begin{aligned}
		\lambda_{\boldsymbol{k}}&=\frac{A_{j}(k)}{\sqrt{1+\epsilon^{2}}}\sqrt{\boldsymbol{p}_{x}^{2}+\epsilon^{2}\boldsymbol{p}_{y}^{2}}\delta(\boldsymbol{k}-\boldsymbol{k}_{0}),\quad
		\theta_{\boldsymbol{k}}=\phi_{\text{cep}}+ \Lambda\arctan(\epsilon\tan(\varphi_{p})),\\
		\rho_{\boldsymbol{k}}&=\frac{A_{j}(\boldsymbol{k})}{\sqrt{1+\epsilon^{2}}}\frac{\sqrt{\boldsymbol{p}_{x}^{2}+\epsilon^{2}\boldsymbol{p}_{y}^{2}}}{\eta_{j}(\boldsymbol{k})}\delta(\boldsymbol{k}-\boldsymbol{k}_{0}),\quad
		\Delta^{+}_{\boldsymbol{k},\boldsymbol{k}'}=\frac{A_{j}(\boldsymbol{k})A_{j}((\boldsymbol{k}')}{4}\frac{1-\epsilon^{2}}{1+\epsilon^{2}}\delta(\boldsymbol{k}-\boldsymbol{k}_{0})\delta(\boldsymbol{k}'-\boldsymbol{k}_{0}),\\
		\Delta^{-}_{\boldsymbol{k},\boldsymbol{k}'}&=\frac{A_{j}(\boldsymbol{k})A_{j}((\boldsymbol{k}')}{4}\delta(\boldsymbol{k}-\boldsymbol{k}_{0})\delta(\boldsymbol{k}'-\boldsymbol{k}_{0}),\quad
		\theta^{+}_{\boldsymbol{k}}=2\phi_{\text{cep}}\;\text{and}\;\theta^{-}_{\boldsymbol{k}}=0,\quad
		\sigma_{\boldsymbol{k},\boldsymbol{k}'}=0,\; \xi_{\boldsymbol{k},\boldsymbol{k}'}=\phi_{\text{cep}},\\
		\alpha^{\pm}_{\boldsymbol{k},\boldsymbol{k}'}&=\frac{A_{j}(\boldsymbol{k})A_{j}((\boldsymbol{k}')}{4}\frac{1\mp\epsilon^{2}}{1+\epsilon^{2}}\frac{1}{\eta_{j}(\boldsymbol{k})\pm\eta_{j}(\boldsymbol{k'})}\delta(\boldsymbol{k}-\boldsymbol{k}_{0})\delta(\boldsymbol{k}'-\boldsymbol{k}_{0}),
		\label{Eq:B4}
	\end{aligned}
\end{equation}
where $\eta_{\boldsymbol{k}}=\boldsymbol{p}.\boldsymbol{k}-\omega_{k}$ and $\omega_{k}=kc$. To evaluate the phase, $\Gamma(\boldsymbol{r},t)$, we can use Eq. (\ref{eq:10}) and the vector potential from Eq. (\ref{eq:13}). By inserting (\ref{Eq:B4}) into Eq. (\ref{eq:10}), we obtain the desired result, given by
	\begin{equation}
		\begin{aligned}
			\Gamma(\boldsymbol{r},t) = & \int d^{3}\boldsymbol{k} \frac{A_{j}(\boldsymbol{k})}{\sqrt{1+\epsilon^{2}}}\frac{\sqrt{\boldsymbol{p}_{x}^{2}+\epsilon^{2}\boldsymbol{p}_{y}^{2}}}{\eta_{j}(\boldsymbol{k})}\sin(\text{u}_j+\phi_{\text{cep}}+ \Lambda\arctan(\epsilon\tan(\varphi_{p})))\delta(\boldsymbol{k}-\boldsymbol{k}_{0})\\
			& + \int d^{3}\boldsymbol{k} \int d^{3}\boldsymbol{k}' \frac{A_{j}(\boldsymbol{k})A_{j}((\boldsymbol{k}')}{4}\frac{1-\epsilon^{2}}{1+\epsilon^{2}}\frac{1}{\eta_{j}(\boldsymbol{k})+\eta_{j}(\boldsymbol{k'})}\sin(\text{u}_j+\text{u}'_{j}+2\phi_{\text{cep}})\delta(\boldsymbol{k}-\boldsymbol{k}_{0})\delta(\boldsymbol{k}'-\boldsymbol{k}_{0})\\
			& + \int d^{3}\boldsymbol{k} \int d^{3}\boldsymbol{k}' \frac{A_{j}(\boldsymbol{k})A_{j}((\boldsymbol{k}')}{4}\frac{1}{\eta_{j}(\boldsymbol{k})-\eta_{j}(\boldsymbol{k'})}\sin(\text{u}_j+\text{u}'_{j})\delta(\boldsymbol{k}-\boldsymbol{k}_{0})\delta(\boldsymbol{k}'-\boldsymbol{k}_{0}).
		\end{aligned}
		\label{Eq:B5}
	\end{equation}
We can utilize the definition of the delta function to simplify the above equation, resulting in Eq. (\ref{eq:14}). Additionally, the third integral in Eq. (\ref{eq:10}) vanishes due to the fact that the k-vectors are parallel to the direction of laser propagation (z-axis), causing the value of $\sigma_{\boldsymbol{k},\boldsymbol{k}'}$ to be zero.
\end{widetext}

\bibliography{SFA-Bib}

\end{document}